\documentclass{iau}
\usepackage{graphicx}
\def\HI{H{\,\small I}}

\title[Jet-driven outflows of  \HI\ and molecular gas] 
{Radio jets clearing the way through galaxies: the view from \HI\ and molecular gas}

\author[R. Morganti]   
{
Raffaella Morganti
}

\affiliation{
ASTRON, the Netherlands Institute for Radio Astronomy, Postbus 2, 7990 AA, Dwingeloo, The Netherlands \\
and \\
Kapteyn Astronomical Institute, University of Groningen, Postbus 800, \\
9700 AV Groningen, The Netherlands \\
email: {\tt morganti@astron.nl} \\
}

\pubyear{2014}
\volume{313} 
\pagerange{xx--xx}
\jname{Extragalactic jets from every angle}
\editors{F. Massaro, C.C. Cheung, E. Lopez, A. Siemiginowska, eds.}

\begin{document}

\maketitle

\begin{abstract}
Massive gas outflows are considered a key component in the process of galaxy formation and evolution. Because of this, they are the topic of many studies aimed at learning more about their occurrence, location and physical conditions as well as the mechanism(s) at their origin.  
This contribution presents recent results on two of the best examples of jet-driven outflows traced by cold and molecular gas.  Thanks to high-spatial resolution observations, we have been able to locate the region where the outflow occurs. This appears to be coincident with bright radio features and regions where the interaction between radio plasma jet and ISM is known to occur, thus strongly supporting the idea of jet-driven outflows.  We have also imaged the distribution of the outflowing gas. The results clearly show the effect that expanding radio jets and lobes have on the ISM. This appears to be in good  agreement with what predicted from numerical simulations. Furthermore, the results show that cold gas is associated with these powerful phenomena and can be formed - likely via efficient cooling - even after a strong interaction and fast shocks. 
The discovery of similar fast outflows of cold gas in weak radio sources is further increasing the relevance that the effect of the radio plasma can have on the surrounding medium and on the host galaxy.  

\end{abstract}

\firstsection 
\section{Introduction}

The impact of the energy released by an active nucleus (AGN) with the surrounding interstellar medium (ISM) and the consequent production of gas outflows has been known for a long time (e.g. Heckman et al. 1981; Whittle 1985, 1992; Gelderman \& Whittle 1994).
The large energetics connected to these phenomena initially suggested that gaseous outflow would  be traced more naturally by ionized (hot or warm) gas. Indeed, a number of studies have shown that  such outflows, observed  both in X-ray (Reeves et al. 2009; Tombesi et al. 2012, 2014; Fabian 2013) as well as in optical (Nesvadba et al. 2006, 2007, 2008; Alexander et al. 2010; Rosario et al. 2010;  Mullaney et al. 2013; Harrison et al. 2012, 2014; Holt et al. 2009, 2011), are common in AGN. 

However, the discovery that fast and massive outflows can also be traced by cold gas (\HI\ and CO, see e.g. Morganti et al. 2005a,b; Kanekar \& Chengalur 2008; Feruglio et al. 2010; Dasyra \& Combes 2011, 2012; Morganti et al. 2013a,b; Cicone et al. 2014; Garc{\'{\i}}a-Burillo et al. 2014) has challenged our ideas of how exactly the energy released by an AGN may interact with its surroundings. 
One of the reasons why outflows have attracted a lot of attention is that they could play an important role in regulating  the growth of the super-massive black holes and/or the quenching of star formation in early-type galaxies.
Thus, in order to understand if this is the case, we need to have a better insight on the complexity of these structures, build a more complete and realistic view of feedback effects and provide better constraints to theoretical models (see e.g. Fabian 2013, Combes 2014 for reviews).

Different mechanisms have been proposed to accelerate  the gas. Although radiation pressure launching  (wide) winds from the accretion disk interacting and shocking the surrounding medium  (Zubovas \& King, 2012, 2014; Zubovas \& Nayakshin  2014, Costa et al. 2014, Faucher-Gigu{\`e}re \& Quataert 2012 )  is often favorite (Cicone et al. 2014), the role of the radio plasma has also gained interest. 
This mechanism can be particularly relevant in early-type galaxies where up to $\sim 30$\% of the high mass galaxies are radio-loud (Best et al. 2005) and the radio-loud phase is known  to be recurrent. Actually, the high efficiency in the coupling between the radio plasma and the surrounding ISM/IGM and the mechanical power of the radio jets exceeding the synchrotron power (McNamara et al. 2012, B\^irzan et al. 2008,  Cavagnolo et al. 2010), suggest that this mechanism can  be relevant also for relatively weak radio sources. Indeed, a few cases are already  known (see e.g. NGC~1266 Alatalo et al. 2011, Nyland et al. 2013 and NGC~1433 Combes et al. 2013).  The role of radio jets has been also emphasized by the results from numerical simulations (Wagner \& Bicknell 2011, Wagner, Bicknell, Umemura, 2012, G. Bicknell These Proceedings).

While one mechanism  does not exclude the other, it is important to have an overview of the impact that each of them (radiation pressure and mechanical energy from  plasma jets) has on the surrounding ISM.

\begin{figure*}[b]
\begin{center}
\centerline{
\includegraphics[width=4.5cm]{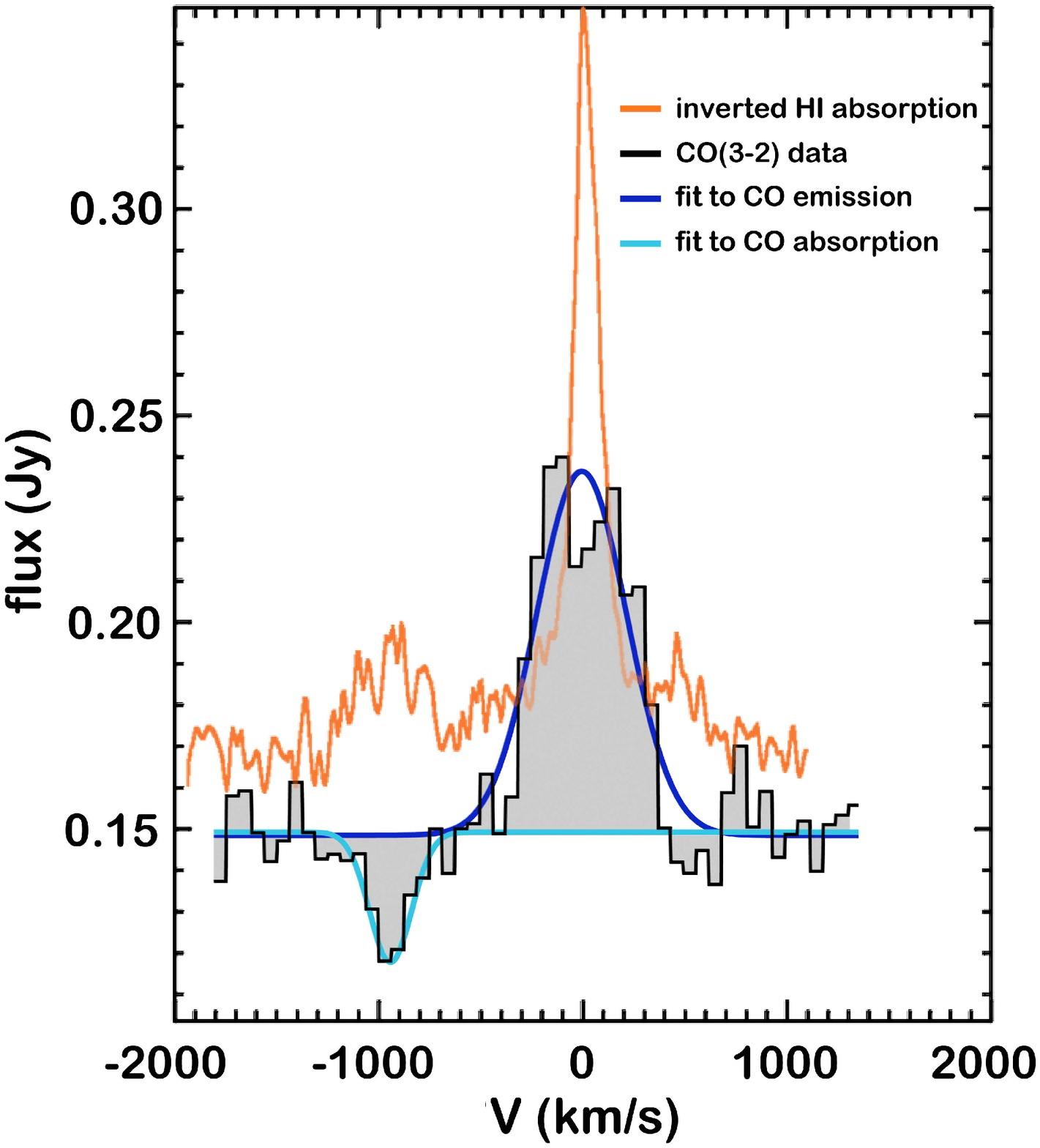}
\includegraphics[width=8cm]{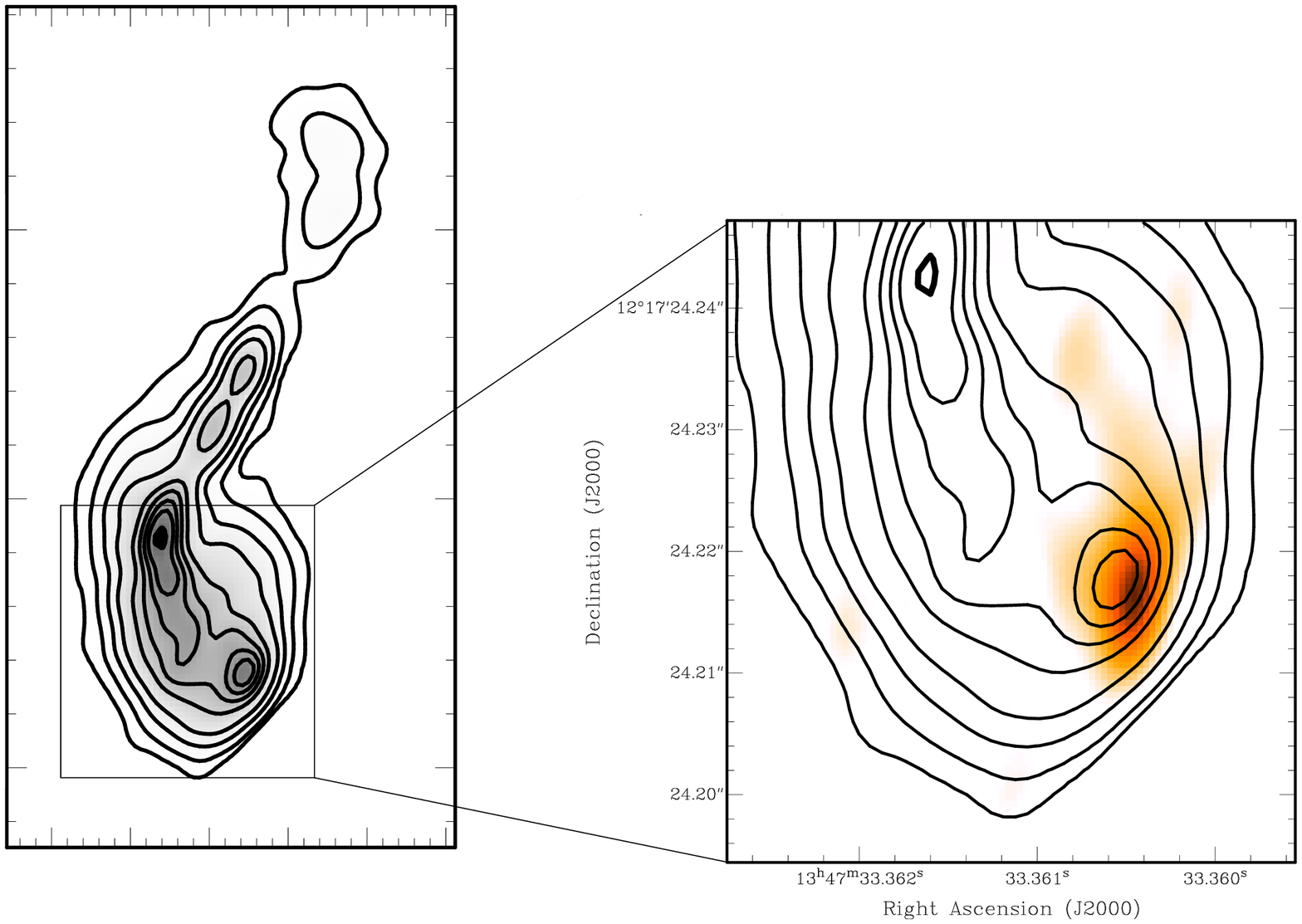} 
}
 \caption{Left: the integrated \HI\ absorption profile is shown superimposed on the CO profile
(taken from Dasyra \& Combes (2012)  with the \HI\ from Morganti et al. (2005a)  inverted for comparison). Middle: radio continuum image of 4C~12.50. Right: zoom-in of the southern lobe of 4C12.50 (contours) with the distribution of the \HI\ detected in absorption indicated in color. }
   \label{fig1}
\end{center}
\end{figure*}

Expanding the number of known outflows of \HI\ and molecular gas in order to provide a larger statistics is the focus of various studies. Observations  of molecular outflows in Ultra Luminous IR Galaxies (combined with known cases of molecular outflows) have been presented by Cicone et al. (2014) showing relations between outflow rates and AGN properties.  In the case of the atomic neutral hydrogen, their occurrence has started to be addressed by a shallow survey (e.g. Ger{\'e}b et al. 2014a,b) that suggests that \HI\ outflows could last only for a fraction of the life of the radio source, i.e. a few Myr up to $10^7$. Thus, they appear to be a temporary phenomena that can, however, be recurrent.

However, the other important direction where progresses need to be made is  to spatially resolve the outflows and trace their distribution and the physical properties of the gas.

Here we summarize the results recently obtained for two interesting radio sources: the young radio galaxy 4C~12.50 and the Seyfert~2 IC~5063. These objects represent two of the most convincing cases of jet-driven outflows traced by \HI\ and molecular gas, thus showing that these phases of the gas can co-exist with energetic processes and shocks. Interestingly, the two sources cover a very broad range of radio power: while 4C~12.50 is a powerful radio source with P$_{1.4} \sim 10^{26}$ W Hz$^{-1}$, IC~5063 is a radio-loud Seyfert galaxy with P$_{1.4} \sim 3 \times 10^{23}$ W Hz$^{-1}$. The latter is  comparable with NGC~1068 but located at the lower edge of the power distribution for radio galaxies. The jet powers of these two objects are, however, not too different and similar to other radio galaxies showing \HI\ outflows (see Guillard et al. 2012 for details).
 
The goal of the work on these objects is to locate the outflows, derive their characteristics and the
physical conditions of the gas in order to understand the impact that radio plasma jets can have.
For a third object, 3C~293, the study of the \HI\ and ionized gas is presented in Mahony et al. (2013) and These Proceedings.

\section{Two cases of jet-driven fast outflows of \HI and molecular gas} 

\subsection{The young, far-IR bright 4C~12.50}

The young radio galaxy 4C~12.50 (PKS 1345+12) is a prime target for AGN feedback studies
as it contains all of the signatures of a recently triggered powerful AGN currently shredding its natal cocoon. 
Holt et al. (2003, 2011) have studied in detail the kinematics of the ionized gas. At the position of the nucleus they observe complex emission line profiles and Gaussian fits to the [OIII] emission lines required three components (narrow, intermediate and broad), the broadest of which has width $\sim 2000$ km s$^{-1}$ (FWHM) and is blueshifted by $\sim 2000$ km s$^{-1}$ with respect to the halo of the galaxy. In \HI, a broad and mostly blueshifted absorption structure was detected by Morganti et al. (2005a).
Interestingly, a very similar profile (shown in Fig. 1, left) was observed for the  molecular gas [CO(1-0) and (3-2)]  by Dasyra \& Combes (2012). Thus, these gaseous components have been interpreted as being part of the same fast outflow of cold gas.

The signature of what is at the origin of such  fast outflows was found  by tracing the \HI\ outflowing gas down to the pc scale using global VLBI observations (Morganti et al. 2013). Fig. 1 (right)  shows the location of the blueshifted absorption as traced by the VLBI data. 
Blueshifted ($\sim1000$ km s$^{-1}$) \HI\ absorption was detected from two components: a compact ($<10$ pc) cloud located at the end of the southern radio jet - about 100 pc from the core - and a diffuse trail of \HI\ observed against and around the southern radio lobe.  The unresolved cloud was estimated to have a column density of N$_{\rm \HI} = 4.6 \times10^{21}$ cm$^{-2}$ (assuming the temperature T$_{spin} = 100$ K, thus representing a lower limit). The derived mass of the cloud is M$_{\rm \HI} = 600$ M$_{\odot}$, reaching up to  M$_{\rm \HI} =1.6 \times 10^4$ M$_{\odot}$ if the extended part is included. The average densities derived for the compact clouds range between 150 cm$^{-3}$ and 300 cm$^{-3}$. 

\begin{figure*}[b]
\begin{center}
\centerline{
\includegraphics[width=8cm]{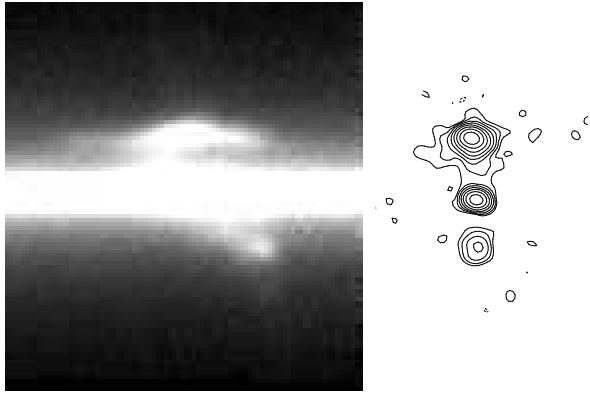} 
}
 \caption{Left: H$_2$ 2.2 $\mu$m ISAAC spectrum taken along the radio axis of IC~5063 (Tadhunter et al. 2014) Right: radio image of IC~5063 obtained from the ATCA (Morganti et al. 2007). The co-spatial location of the broad (FWZI $\sim 1200$ km s$^{-1}$) H$_2$ line and the bright radio lobe ($\sim 0.5$ kpc from the radio core) is evident.}
   \label{fig2}
\end{center}
\end{figure*}

\subsection{The molecular outflow in the radio-loud Seyfert galaxy IC~5063}

The Seyfert 2 galaxy IC~5063 has a  lower radio power (P$_{\rm 1.4} = 3 \times10^{23}$ W Hz$^{-1}$) than 4C~12.50. However, like 4C~12.50, it  shows prominent outflows of different phases of the gas, from \HI\ to ionized gas, and they have been presented and discussed  in Morganti et al. (2007 and references therein). These components of the outflows have been identified associated with the brighter radio lobe, about 0.5 kpc from the nucleus. In the case of the \HI, this was done with single-baseline VLBI observation (Oosterloo et al. 2000)  and, therefore, the distribution of the gas could not be imaged as in the case of 4C~12.50. 

The conditions of the molecular gas in IC~5063 have been first explored in CO(2-1) using APEX. From these observations, an outflow of $\sim  10^7$ M$_\odot$ has been reported by Morganti et al. (2013b).  However, the low spatial resolution of these observations did not allow to trace the location of the outflowing gas.
The outflow of warm molecular gas has been further explored with observations of the  H$_2$ 1-0 S(1) at  2.128 $\mu$m line using ISAAC at the VLT. Tadhunter et al. (2014) reported the identification of the location of the outflow, corresponding to the brighter radio lobe (see Fig. 2) about 0.5 kpc from the nucleus. This confirmed, for the first time, the possibility of having jet-driven  molecular outflows.  The molecular gas shows a broad, complex profile of the H$_2$ emission line with a full width at zero intensity of FWZI $\sim 1200$ km s$^{-1}$. At the location of the brighter radio lobe, the profile is clearly broader than that of the nucleus or of the eastern radio lobe (see Fig. 2). A temperature of $\sim 1900$ K and a molecular hydrogen mass of M$_{\rm H_2} \sim 8 \times 10^2$ M$_\odot$ for the western outflow region were derived (see Tadhunter et al. 2014 for details).

However, even more spectacular appears to be the distribution and kinematics of cold molecular gas traced by CO(2-1) when observed at the high spatial resolution and sensitivity of ALMA (Morganti et al. in prep).  The observations, using the most extended configuration ($\sim 0.5$ arcsec resolution),  have allow to resolve the distribution of the molecular gas and locate the outflowing gas by separating it from the regularly rotating component. The observations confirm that the gas with the highest outflowing velocity is located co-spatial with the bright radio lobe. However, there is more to it. Molecular gas with disturbed kinematics is present at all locations along the radio emission (Morganti et al. in prep). At these locations the gas shows  blueshifted and redshifted velocities compared to the regularly rotating component (likely the inner counterpart of the large scale \HI\ disk observed in this galaxy, Oosterloo et al. 2000).  Interestingly, the mass of the disturbed/outflowing component is a few $\times 10^7$ M$_\odot$, thus much larger  than the component of warm molecular gas and also larger than the \HI\ component.  

\section{Origin of the \HI\ and molecular outflows}

In both cases, the extreme kinematics, together with the location  of the outflowing gas, suggest that we are witnessing gas being expelled from the galaxy  as a result of the interaction between the radio jet and dense cloud(s) in the ISM.
The energetics of the radio jets in these sources appear to able to support these outflows, with jet powers more than one order of magnitude larger than the kinetic power of the outflow.

The scenario that appears to explain better our observations is the one in which the relativistic jets are expanding through the clumpy interstellar medium, driving fast shocks into dense molecular clouds embedded in a lower-density medium as suggested in the model of Wagner et al. 2011, 2012.  
Most of the interaction is happening via the large cocoon of disturbed and outflowing gas created around the jet by the interaction of the radio plasma. This cocoon is affecting a large region of the galaxy (see simulations by Wagner et al. 2011, 2012). Direct interaction between the jet and the ISM may occur in some limited regions, e.g. where the jet encounters large, compact  clouds (see e.g. the case of 4C~12.50).
A further support to this scenario is the similarity of the density of the \HI\ clouds detected in 4C~12.50 and those of the clouds in the numerical simulations (Wagner et al. 2011, 2012).
This scenario, together with the larger amount (compared to the other phases of the gas) of cold molecular gas found in the case of IC~5063, supports the possibility of the  atomic and molecular phases being the result of the cooling process of the gas after being warmed up (and possibly ionized) by the passage of a fast shock.

\section{Final considerations}

The results presented emphasize the effects - in some cases surprising - that the interaction between the radio plasma  can have on the surrounding ISM. In the two objects presented, 4C~12.50 and IC~5063,  the location and the physical conditions of the outflowing gas have been derived.  

They show that:
\begin{itemize}
\item outflows of  \HI, molecular and ionized gas co-exist, thus showing that outflows are {\sl truly multiphase};  
\item the location of  the faster outflows is off-nucleus and co-spatial with  bright radio features. However, disturbed outflowing gas is present also along fainter regions of the radio jets/lobes;
\item the most likely scenario describing the observations is a {\sl jet expanding in a clumpy medium}. A combination of two mechanisms can be responsible for the outflows: lateral expansion of the gas pushed by the
jetÕs cocoon and, limited to the brighter radio regions, direct jet/ISM interaction;
\item the gas must be efficiently cooling after the shock produced by the jet.  The cold molecular phase will be the final product of this process, while the warm molecular and \HI\ are the intermediate (and less massive) phases;.
\end{itemize}

The two objects presented here have quite different radio luminosity, with IC~5063 at the lower end of the distribution of radio power for radio galaxies.  The discovery of other cases  (see e.g. NGC~1266 Alatalo et al. 2011, Nyland et al. 2013 and NGC~1433 Combes et al. 2013), where a weak radio source seems to be the only realistic source of  energy for driving a fast outflows of cold gas, further increases the relevance of such mechanism and the importance of studying the occurrence and characteristics.  

\begin{acknowledgements}

The results presented here would not have been obtained without the help of my collaborators. In particular I would like to thank Clive Tadhunter, Tom Oosterloo, Zsolt Paragi, Raymond Oonk, Elizabeth Mahony and Wilfred Frieswijk. 
RM gratefully acknowledge support from the European Research Council under the European UnionÕs Seventh
Framework Programme (FP/2007-2013) /ERC Advanced Grant RADIOLIFE-320745.

\end{acknowledgements}


\begin{thebibliography}{}


\bibitem[ala11]{ala11}Alatalo et al.  2011  ApJ 735, 88

\bibitem[]{}  Alexander et al., 2010, MNRAS, 402, 2211

\bibitem[\protect\citeauthoryear{Best et al.}{2005}]{2005MNRAS.362...25B} 
Best P.~N., Kauffmann G., Heckman T.~M., Brinchmann J., et al. 2005, MNRAS, 362, 25 

\bibitem[]{} B\^irzan et al. 2008, ApJ 686, 859; 


\bibitem[\protect\citeauthoryear{Cavagnolo et al.}{2010}]{2010ApJ...720.1066C} Cavagnolo et al., 2010, ApJ, 720, 1066

\bibitem[\protect\citeauthoryear{Cicone et al.}{2014}]{2014A&A...562A..21C} Cicone  et al., 2014, A\&A, 562, A21 

\bibitem[\protect\citeauthoryear{Combes}{2014}]{2014arXiv1408.1591C} Combes 
F., 2014,  Proceedings of IAU Symp-309, ed. B.L. Ziegler et al., arXiv:1408.1591  

\bibitem[\protect\citeauthoryear{Combes et  al.}{2013}]{2013A&A...558A.124C} Combes F., et al., 2013, A\&A, 558, A124 

\bibitem[\protect\citeauthoryear{Costa, Sijacki, 
\& Haehnelt}{2014}]{2014MNRAS.444.2355C} Costa T., Sijacki D., Haehnelt M.~G., 2014, MNRAS, 444, 2355 

\bibitem[\protect\citeauthoryear{Dasyra \& Combes}{2011}]{2011A&A...533L..10D} Dasyra K.~M., Combes F., 2011, A\&A, 533, L10 

\bibitem[]{} Dasyra K.~M., Combes F. 2012 A\&A 541, L7


\bibitem[]{} Fabian, A. 2013,  Ann.Rev.Astron.Astrophys. 50, 455

\bibitem[\protect\citeauthoryear{Faucher-Gigu{\`e}re 
\& Quataert}{2012}]{2012MNRAS.425..605F} Faucher-Gigu{\`e}re C.-A., Quataert E., 2012, MNRAS, 425, 605 


\bibitem[\protect\citeauthoryear{Feruglio et 
al.}{2010}]{2010A&A...518L.155F} Feruglio et al., 2010, A\&A, 518, L155; 


\bibitem[\protect\citeauthoryear{Garc{\'{\i}}a-Burillo et 
al.}{2014}]{2014A&A...567A.125G} Garc{\'{\i}}a-Burillo S., et al., 2014, A\&A, 567, 125 

\bibitem[]{}  Gelderman R., Whittle M., 1994, APJS, 91, 491

\bibitem[\protect\citeauthoryear{Ger{\'e}b, Morganti, 
\& Oosterloo}{2014}]{2014A&A...569A..35G} Ger{\'e}b K., Morganti R., Oosterloo T.~A., 2014a, A\&A, 569, AA35

\bibitem[]{} Ger{\'e}b K., Maccagni F., Morganti  R., Oosterloo T. et al. 2014b, A\&A in press (arXiv:1411.0361)

\bibitem[\protect\citeauthoryear{Guillard et  al.}{2012}]{2012ApJ...747...95G} Guillard P., et al., 2012, ApJ, 747, 95 

\bibitem[\protect\citeauthoryear{Harrison et  al.}{2012}]{2012MNRAS.426.1073H} Harrison C.~M., et al., 2012, MNRAS, 426, 1073

\bibitem[\protect\citeauthoryear{Harrison et 
al.}{2014}]{2014MNRAS.441.3306H} Harrison C.~M., Alexander D.~M., Mullaney 
J.~R., Swinbank A.~M., 2014, MNRAS, 441, 3306 

\bibitem[]{}  Heckman T. M., Miley G. K., van Breugel W. J. M., Butcher H. R.,
1981, ApJ, 247, 403

\bibitem[\protect\citeauthoryear{Holt et al.}{2011}]{2011MNRAS.410.1527H} 
Holt J., Tadhunter C.~N., Morganti R., Emonts B.~H.~C., 2011, MNRAS, 410, 
1527 

\bibitem[\protect\citeauthoryear{Holt, Tadhunter,  \& Morganti}{2009}]{2009MNRAS.400..589H} Holt J., Tadhunter C.~N., Morganti R., 2009, MNRAS, 400, 589 

\bibitem[\protect\citeauthoryear{Holt, Tadhunter, 
\& Morganti}{2003}]{2003MNRAS.342..227H} Holt J., Tadhunter C.~N., Morganti R., 2003, MNRAS, 342, 227 

\bibitem[{Kanekar \& Chengalur(2008)}]{kanekar08} Kanekar, N. \& Chengalur, J.~N. 2008, MNRAS, 384, 6L

\bibitem[]{}Mahony E., Morganti R., Emonts B., Oosterloo T., Tadhunter C., 2013, MNRAS, 435, L58

\bibitem[\protect\citeauthoryear{McNamara 
\& Nulsen}{2012}]{2012NJPh...14e5023M} McNamara B.~R., Nulsen P.~E.~J., 2012, NJPh, 14, 055023 

\bibitem[]{}  Morganti, R., Fogasy, J., Paragi, Z., Oosterloo, T., Orienti, M. 2013a,  Science 341, 1082;

\bibitem[]{} Morganti R., Frieswijk W., Oonk R., Oosterloo T., Tadhunter C. 2013b,  A\&A  552, L4;

\bibitem[]{}Morganti, R., Tadhunter, C. N., Oosterloo, T. A. 2005a A\&A 444, L9; 

\bibitem[]{}Morganti et al. 2005b, A\&A, 439, 521;  

\bibitem[\protect\citeauthoryear{Morganti et al.}{2007}]{morganti07} Morganti R., Holt J., Saripalli L., Oosterloo T.~A., Tadhunter C.~N., 2007, A\&A, 476, 735 

\bibitem[\protect\citeauthoryear{Mullaney et al.}{2013}]{2013MNRAS.433..622M} Mullaney J.~R., Alexander D.~M., Fine S., 
Goulding A.~D. et al., 2013, MNRAS, 433, 622 




\bibitem[\protect\citeauthoryear{Nesvadba et  al.}{2008}]{2008A&A...491..407N} Nesvadba N.~P.~H., Lehnert M.~D., De Breuck C. et al.,  A\&A, 491, 407 

\bibitem[\protect\citeauthoryear{Nesvadba et  al.}{2007}]{2007A&A...475..145N} Nesvadba N.~P.~H., Lehnert M.~D., De Breuck C. et al., 2007, A\&A, 475, 145 

\bibitem[\protect\citeauthoryear{Nesvadba et 
al.}{2006}]{2006ApJ...650..693N} Nesvadba N.~P.~H., Lehnert M.~D., Eisenhauer F., Gilbert A. et al., 2006, ApJ, 650, 693 

\bibitem[\protect\citeauthoryear{Nyland et al.}{2013}]{2013ApJ...779..173N} 
Nyland K., et al., 2013, ApJ, 779, 173 

\bibitem[\protect\citeauthoryear{Oosterloo et 
al.}{2000}]{2000AJ....119.2085O} Oosterloo T., Morganti R., Tzioumis A., 
Reynolds J., King E. et al., 2000, AJ, 119, 2085 

\bibitem[]{} Reeves, J. N., et al. 2009, ApJ, 701, 493

\bibitem[]{} Rosario D. J., Shields G. A., Taylor G. B., Salviander S., Smith
K. L., 2010, ApJ, 716, 131

\bibitem[\protect\citeauthoryear{Tadhunter et 
al.}{2014}]{2014Natur.511..440T} Tadhunter C., Morganti R., Rose M., Oonk 
J.~B.~R., Oosterloo T., 2014, Natur, 511, 440;


\bibitem[\protect\citeauthoryear{Tombesi et  al.}{2014}]{2014MNRAS.443.2154T} Tombesi F., Tazaki F., Mushotzky R.~F., 
Ueda Y., Cappi M., et al. 2014, MNRAS, 
443, 2154 

\bibitem[\protect\citeauthoryear{Tombesi et  al.}{2012}]{2012MNRAS.422L...1T} Tombesi F., Cappi M., Reeves J.~N., Braito  V., 2012, MNRAS, 422, L1 

\bibitem[\protect\citeauthoryear{Zubovas 
\& Nayakshin}{2014}]{2014MNRAS.440.2625Z} Zubovas K., Nayakshin S., 2014, MNRAS, 440, 2625 

\bibitem[\protect\citeauthoryear{Zubovas 
\& King}{2014}]{2014MNRAS.439..400Z} Zubovas K., King A., 2014, MNRAS, 439, 400 

\bibitem[\protect\citeauthoryear{Zubovas 
\& King}{2012}]{2012ApJ...745L..34Z} Zubovas K., King A., 2012, ApJ, 745, L34 

\bibitem[\protect\citeauthoryear{Wagner, Bicknell, 
\& Umemura}{2012}]{2012ApJ...757..136W} Wagner A.~Y., Bicknell G.~V., Umemura M., 2012, ApJ, 757, 136 

\bibitem[\protect\citeauthoryear{Wagner 
\& Bicknell}{2011}]{2011ApJ...728...29W} Wagner A.~Y., Bicknell G.~V., 2011, ApJ, 728, 29 

\bibitem[]{} Whittle M., 1985, MNRAS, 213, 1

\bibitem[]{} Whittle M., 1992, ApJ, 387, 109


\end{thebibliography}
\end{document}